\documentclass[11pt,twoside]{article}

\usepackage{asp2006}
\usepackage{epsf}
\usepackage{psfig}
\usepackage{lscape}
\usepackage{graphicx}

\begin{document}
\title{Massive stars and high-energy neutrinos}   
\author{Gustavo E. Romero$^{1,2}$}   
\affil{\affil{%
(1) Instituto Argentino de Radioastronom\'\i a, CCT La Plata-CONICET, C.C. 5, 1894 Villa Elisa, Argentina; romero@iar-conicet.gov.ar\\
(2) Facultad de Ciencias Astron\'omicas y Geof\'\i sicas, Paseo del Bosque s/n, 1900 La Plata, Argentina; romero@fcaglp.unlp.edu.ar\\}}    

\begin{abstract}

Massive stars have been associated with the production of high-energy
neutrinos since the early claims of detection of very high-energy gamma
rays from Cygnus X-3 in the 1970s and early 1980s. Although such claims are
now discredited, many thoretical models were developed predicting significant
neutrino fluxes from binary systems with massive stars. With the discovery of
microquasars, new, improved models appeared. The large neutrino telescopes
currently under construction (IceCube, Antares) and the detection of gamma-ray
sources of likely hadronic origin associated with massive binaries and star-forming
regions make the prospects for high-energy neutrino astronomy quite
promising. In this paper I will review the basic features of neutrino production in
stellar systems and I will discuss the physical implications of a positive neutrino
detection from such systems for our view of the stellar evolution and comsic ray
origin.
\end{abstract}

\section{Introduction}  

The source of energy in main sequence stars are thermonuclear reactions, hence neutrino emission is a natural byproduct of the processes that keep the stellar equilibrium. Solar neutrinos, with energies $\sim 400$ keV, have been detected from the Sun. These neutrinos are the result of the proton-proton chain reaction:
$$ 4p+2e\rightarrow {\rm He} + 2\nu_{e}.$$ There are also several other significant production mechanisms, with energies up to 10 MeV (see Bellerive 2004, for a review on solar neutrino experiments). In the case of high-mass stars with $M\geq 8$ $M_{\odot}$, carbon ignites in the core and, depending on the mass, further stages of nuclear burning occur. A sufficiently massive star has an ``onion'' structure, with iron at the core and successive layers of nickel, silicon, oxygen, neon, carbon, helium, and hydrogen, with different reactions at different levels. Neutrinos are produced in reactions of the type $$e^{+}(Z,\;A)\rightarrow (Z-1,\; A)+ \nu (Z,\;A) + e + \bar{\nu}.$$ These neutrinos can carry most of the energy output of a high-mass star. The cross section for neutrinos is very small and scales with the square of the neutrino energy. Hence, neutrinos produced by thermonuclear reactions in high-mass stars are undetectable from the earth. 

High-energy neutrinos, with energies above 1 TeV, have cross sections that make possible their detection with telescopes of cubic-kilometer scale, like IceCube and KM3NeT (see Spiering 2008 for a recent review). The generation of such neutrinos requires sources capable to accelerate protons up to very high energies. Neutrinos then result from the decay of charged pions created in proton-proton or proton-photon interactions. The same interactions also produce neutral pions, which decay into gamma rays. Hence, neutrino and gamma-ray astronomy are intimately interconnected. Local absorption of gamma rays due to compactness or other ambient photon fields in the sources, can prevent an effective detection of electromagnetic signals because of photon annihilation. Hence, dark neutrino sources might also exist. In any case, neutrinos can be used to probe the conditions in the innermost regions of cosmic ray accelerators.

Since the early claims of detection of very high-energy gamma rays from Cygnus X-3 in the 1970s and early 1980s, models for high-energy proton interactions in the vicinity of massive, hot stars have been developed. Cygnus X-3 has been discredited as a very high-energy photon source long ago (see Weekes 2003 for a summary of the controversy around this system), but the models then developed were the seed for new theoretical predictions about neutrino production in galactic objects that will be tested by the forthcoming generation of neutrino telescopes. 

Many of these models explore the possibility that one or more massive stars might play an important role in the generation of the neutrinos. Here, we shall review some of such models and the prospects for detection of high-energy neutrinos from stellar systems.

\section{Models involving compact objects}

\begin{figure}[!ht]
\begin{center}
\includegraphics[width=10.5cm]{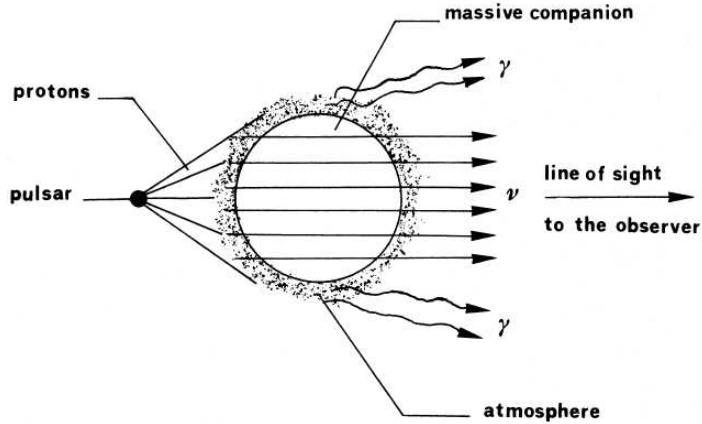}\label{Bere}
\end{center}
\caption{A pulsar in a close orbit around a massive star. The star is irradiated by the relativistic pulsar wind, producing gamma rays and neutrinos. From Berezinsky et al. (1986).}
\end{figure}

The pioneer models originally developed for Cygnus X-3 (Vestrand \& Eichler 1982; Kolb, Turner, \& Walker 1985; Berezinsky, Castagnoli,  \& Galeotti 1986) all adopted the idea that a young pulsar might irradiate with relativistic protons the massive star in a binary system. Then, gamma rays would be produced by interactions between relativistic protons in the pulsar wind and cold (thermal) protons of the outer atmosphere. In this context, neutrinos result from the direct proton impact onto the star, where the gamma rays must be absorbed. The star is essentially transparent to the neutrinos, which could reach the earth to be detected by large arrays of photo-multipliers, located underwater in the deep sea. The light to be detected is the Cherenkov light produced by a locally created muon in the weak interaction of a cosmic neutrino with a local atom. The basic astrophysical scenario is shown in Figure \ref{Bere}, taken from Berezinsky et al.'s (1986) paper.    

A similar scenario was presented by Aharonian \& Atoyan (1996), who considered the impact of a proton beam produced by a pulsar with clouds ejected by a massive star. 

After EGRET instrument onboard the Compton satellite detected many sources on the galactic plane (Romero, Benaglia, \& Torres 1999) and it was established that many of them were variable (e.g. Torres, Pessah, \& Romero 2001), it resulted natural to suggest that some of these sources might be microquasars (Romero et al. 2004; Bosch-Ramon, Romero, \& Paredes 2005). A microquasar is a binary system formed by a donor star and an accreting compact object. These systems present relativistic jets with non-thermal emission (e.g. Mirabel \& Rodr{\'\i}guez 1999). If the donor star is an early-type, hot star, the microquasar is called a `high-mass' microquasar.

Hadronic models for jet interactions with the winds of massive stars were developed in recent years (Romero et al. 2003; Romero, Christiansen, \& Orellana 2005; Christiansen, Orellana, \& Romero 2006). These papers included predictions of the neutrino output from the $pp$ interactions between relativistic protons from the jet and thermal protons from the stellar wind. Possible variants for the interaction mechanism were briefly discussed by Aharonian et al. (2006). Perucho \& Bosch-Ramon (2008) have implemented numerical simulations of the hydrodynamic interaction of a microquasar jet with a stellar wind. Their results shed light on the jet stability in various scenarios and the level of mixing that can be expected between jet and wind. More recently, Owocki et al. (2009) have studied, from a statistical point of view, the effects of clumped structure in the stellar wind upon the gamma-ray variability of microquarsars. Individual flares produced by jet-clump interactions have been studied by Romero et al. (2008) and, with more detail, by Araudo et al. (2009). 

Among the high-mass microquasars for which detailed predictions for the neutrino flux are available, perhaps the most promising seems to be SS 433, a system with particularly powerful jets and clear evidence for the presence of iron nuclei in its jets. Recent time-dependent calculations of neutrino production in the jets and jet-star interaction show upper limits below the current sensitivity of the already deployed strings in IceCube, but within the scope of the full instrument (Reynoso, Romero, \& Christiansen 2008). 

A different scenario for neutrino production in systems where the primary star is a Wolf-Rayet has been developed by Bednarek (2007). In this model, nuclei from the stellar wind are photo-disintegrated by the jet and some of the neutrons released in the process impact onto the accretion disk and the star, producing neutrinos. The flux, however, is not expected to be very high. A stronger neutrino source might result from the direct impact of the jet onto the star, a case studied in detail by Romero \& Orellana (2005). Such a source, however, would have a periodic signal with a duty cycle that will depend on the orbital parameters. An additional effect of the jet-star interaction is the induced nucleosynthesis in the stellar atmosphere (Butt et al. 2003).

\begin{figure}[!ht]
\begin{center}
\includegraphics[width=8.5cm]{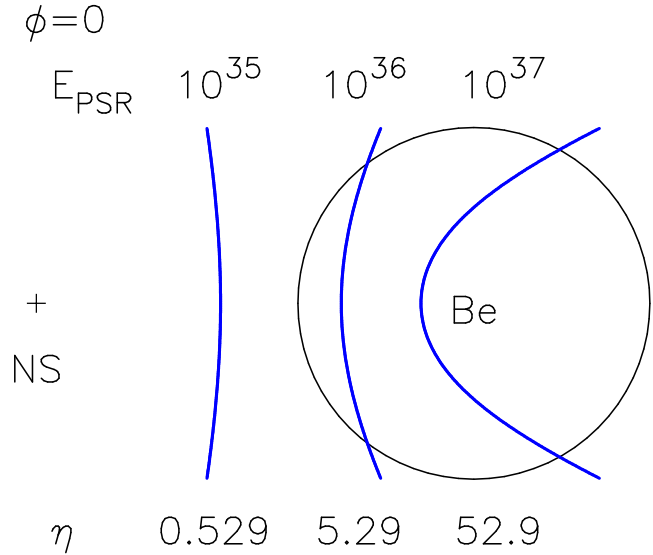}\label{shocks}
\end{center}
\caption{Different interaction surfaces in a pulsar-Be star colliding wind system. The surfaces are calculated using Eq. \ref{eq:eta} for the parameters given in the text during the periastron passage ($\phi=0$). Notice that in the case of strong pulsar winds the interaction surface warps around the star.}
\end{figure}

If the compact object in the massive binary system is a young neutron star, the rotation-driven relativistic wind from the pulsar will interact with the stellar wind or the circumstellar disk in the case of Be stars . If the pulsar wind is baryon-loaded, neutrinos will be produced through $pp$ collisions (Neronov \& Ribordy 2009). In Figure \ref{shocks} we show the wind-wind interaction surface for different cases of pulsar power (the stellar wind corresponds to that of a Be0V star). The key parameter determining the shape of the interaction surface
of the colliding winds is the ratio of wind momentum fluxes, given by 
\begin{equation}
   \eta = \frac{\dot{E}_{\rm PSR}}{\dot{M}_{\rm Be} V_{\rm Be} c}
   \label{eq:eta}
\end{equation}
where $V_{\rm Be}$ and $\dot{M}_{\rm Be}$ are the velocity
and mass loss rate of the Be wind, and  
$\dot{E}_{\rm PSR}$ is the power of the pulsar wind.
For the assumed star (similar to that present in LSI +61 303), we have  $V_{\rm Be}=10^{3}\,{\rm km\,s}^{-1}$ and 
$\dot{M}_{\rm Be}=10^{-8} M_{\odot}\,{\rm yr}^{-1}$. The different curves in the figure corresponds to different pulsar spin-down luminosities in units of erg s$^{-1}$. 

In this simple 2D model that ignores orbital  motion, the interaction surface
approaches a cone with opening half-angle that can be derived from Antokhin et al. (2004):
\begin{equation}
    \theta = 180^{\circ} \, {\eta \over 1 + \eta } \, .
    \label{eq:opang}
\end{equation}

More realistic estimates of the colliding wind surface should take into account the effect of instabilities, radiation pressure, orbital motion, tidal effects, etc. In Figure 3. we show the result of an 3D SPH simulation of colliding winds for the system LSI +61 303, with a pulsar spin-down luminosity of $10^{36}$ erg s$^{-1}$ (Romero et al. 2007). Notice the complexity of the resulting structures, subject to Rayleigh-Taylor and Kelvin-Helmholtz instabilities.  

\begin{figure}[!ht]
\begin{center}
\includegraphics[width=8.5cm]{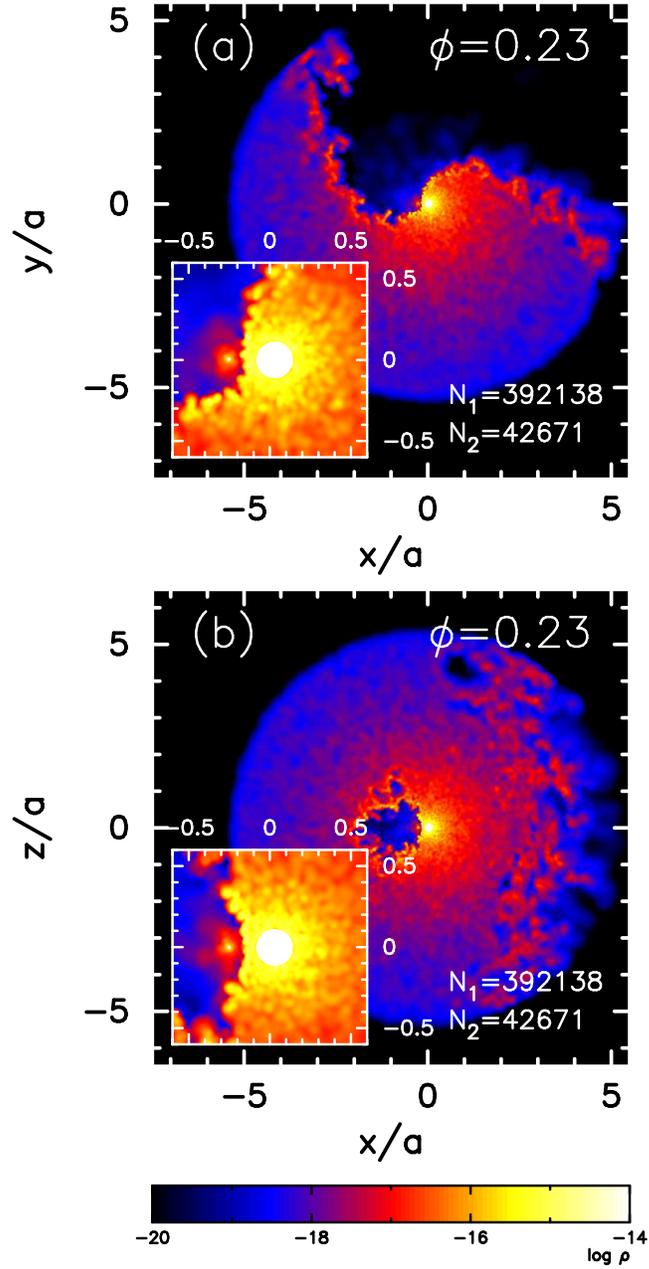}\label{LSI}
\end{center}
\caption{Results of a 3D SPH simulation of colliding winds in a system with a Be0V star and a pulsar with a spin-down luminosity of $10^{36}$ erg s$^{-1}$. The upper panel shows the orbital plane, whereas the lower panel shows the edge-on image. The snapshots are taken at the periastron passage of the pulsar. From Romero et al. (2007).}
\end{figure}

\section{Colliding wind models}

Colliding wind models for systems formed by two massive stars have been extensively studied. The detection of non-thermal radio emission from such systems (see the recent reviews by De Becker 2007 and Benaglia 2009) clearly indicate that colliding wind binaries can accelerate electrons up to high energies. Following the seminal paper by Eichler \& Usov (1993), several authors pointed out that binaries like WR 140, WR 146, WR 147, Cygnus OB2 No.5, etc, might be gamma-ray sources as well (Romero et al. 1999, Benaglia et al. 2001, Benaglia \& Romero 2003). More complete models have been recently presented by Reimer, Pohl, \& Reimer (2006) and Pittard \& Dougherty (2006). All these works conclude that the dominant gamma-ray emission in these systems should be inverse Compton up-scattering of stellar photons (see also Pittard 2009). However, protons could be accelerated in the colliding wind region as well. The cooling time for these protons, dominated by inelastic $pp$ collisions, is $$	[t^{(pp)}_{p}]^{-1}=\frac{1}{E_{p}}\frac{dE_{p}}{dt}=n_{p}c \sigma_{pp} K^{(pp)},$$ where $E_{p}$ is the proton energy, $n_{p}$ the density of thermal protons, $K^{(pp)}\approx 0.5$ the inelasticity (fraction of proton energy lost per interaction) and $\sigma_{pp}\sim 35$ mb is the $pp$ cross section at GeV energies. For a typical density of $n_{p}\sim 10^7$ cm$^{-3}$, $t^{(pp)}_{p}\sim 2\times 10^{8}$ s. The advection time is, roughly, $t^{(\rm adv)}_{p}\sim x/v_{\infty}$, where $v_{\infty}$ is the final wind velocity. Using values from WR 140 (Pittard \& Dougherty 2006), $t^{(\rm adv)}_{p}\sim 10^{6}$ s. Hence, protons can leave the system without significant radiative losses. These protons will diffuse into the interstellar medium. Since massive binaries are usually located in open clusters and OB associations, the protons could reach a nearby molecular cloud producing an extended gamma-ray and neutrino source. Such a scenario has been studied by Benaglia et al. (2005) in connection to the system WR 21a. The total kinetic power in the wind-wind collision is of the order of $\sim 10^{36}$ erg s$^{-1}$.  If $\sim 10$\% of this power goes to relativistic particles (mainly protons), then these can be stored in a nearby cloud yielding a gamma-ray and neutrino source detectable by the full IceCube telescope through long integrations.

\section{Neutrino production in the collapse of very massive stars and gamma-ray bursts}

\begin{figure}[!ht]
\begin{center}
\includegraphics[width=9.5cm]{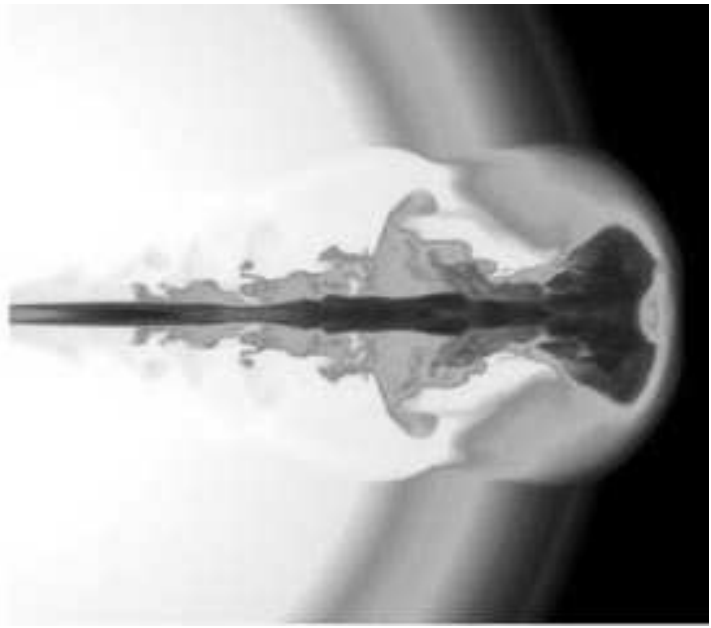}\label{Woosley}
\end{center}
\caption{Simulation of a relativistic jet erupting from the interior of a collapsar. Adapted from Zhang et al. (2006).}
\end{figure}

Very massive stars with high angular momentum may collapse directly to a black hole (Woosley 1993). An hyper-dense accretion disk is formed inside the star, and the black hole accretes matter at very high rates ($0.1-1$ $M_{\odot}$ s$^{-1}$). The combined effects of rotation and extremely high magnetic fields launch jets that pave their way through the star outer layers (e.g. Komissarov et al. 2009). Figure \ref{Woosley} shows a simulation of a jet erupting from a collapsing star (Zhang et al. 2004). 

The interaction of relativistic protons in the jet with thermal X-rays from the heated channel created in the star can result in a burst of neutrinos through photo-pion production (M\'esz\'aros \& Waxman 2001). Even if the jet does not break through the stellar envelope, the choked gamma-ray burst could produce a significant neutrino flux. In such a case there would not be an electromagnetic counterpart. If strong magnetic fields are present in the region where the jet uncorks, the photo-meson production can result from the interaction of relativistic protons with synchrotron photons generated by secondary pairs close to the stellar surface (Dermer \& Atoyan 2003, Romero et al., in preparation). Neutrino production via $pp$ interactions has also been discussed in the literature (Dermer \& Atoyan 2003, Pruet 2003). 

Recently, Eiroa \& Romero (2008) have suggested that neutrino microlensing effects can be used to identify choked gamma-ray bursts. The neutrino signal from a jet that cannot erupt from the star should repeat from the same pixel in the sky if a massive object is interposed.  

\section{Neutrinos from star-forming regions}

Collective shocks can lead to efficient particle acceleration inside star-forming regions (e.g., Bykov \& Fleishman, Parizot et al. 2004). The cosmic ray excess in these regions will produce gamma rays and neutrinos as a result of $pp$ collisions with the ambient gas and the winds of massive stars (e.g. Torres et al. 2004, Bednarek 2007). The cosmic ray content of massive molecular clouds can be enhanced by the injection of relativistic protons accelerated at the termination reverse shocks of protostellar jets (Araudo et al. 2007). If the cosmic ray enhancement inside a star-forming region is high enough, then the region might be detectable as a neutrino source. We notice that several OB associations and young open clusters have been detected at high-energy gamma-rays by imaging atmospheric Cherenkov telescopes and satellites (see Romero 2008 and references therein).   

\section{Physical processes affecting the neutrino production}

\begin{figure}[!ht]
\begin{center}
\includegraphics[width=11.5cm]{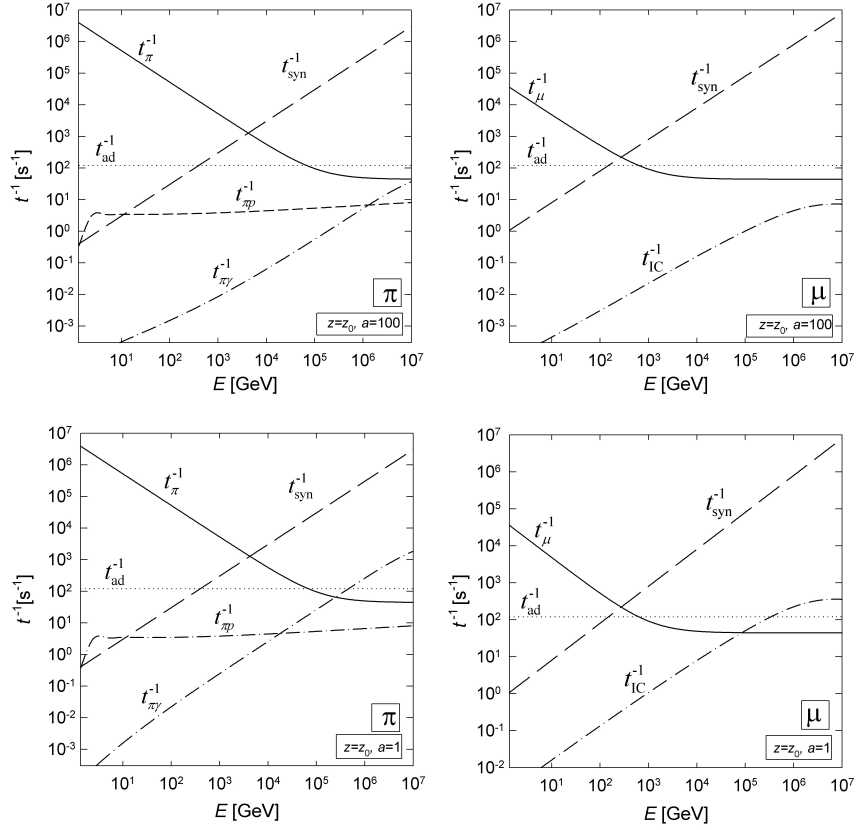}\label{Matias1}
\end{center}
\caption{Losses and decay times for muons and pions at the base of a microquasar jet. From Reynoso \& Romero (2009).}
\end{figure}

\begin{figure}[!ht]
\begin{center}
\includegraphics[width=10.5cm]{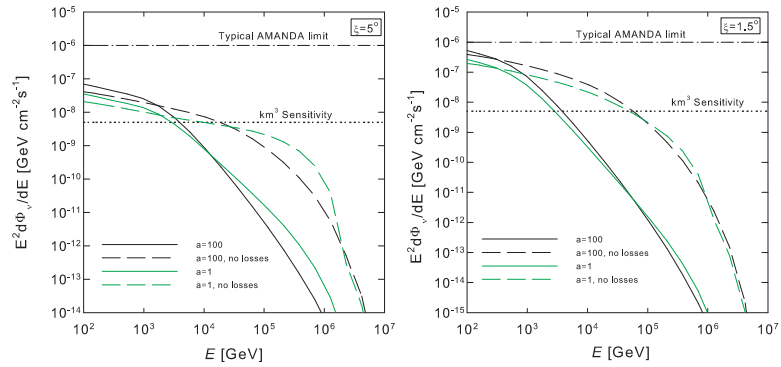}\label{Matias2}
\end{center}
\caption{Modified spectra for neutrino production in a strong magnetic field. From Reynoso \& Romero (2009).}
\end{figure}

If neutrinos are produced in regions with very high magnetic fields, as it seems to be the case of collpsars and the base of jets in microquasars, then muons and pions can significantly cool before decaying, weakening the neutrino signal at the relevant energies for the detectors ($E_{\nu}>1$ TeV). This situation has been discussed in detail by Reynoso \& Romero (2009). In Figure \ref{Matias1} we show the losses and decay time for both muons and pions at the base of the jet of a microquasar. The different values of the parameter $a$ correspond to different proton-to-electron ratios. The magnetic field is $\sim 10^{8}$ G. In Figure \ref{Matias2} the effect of these losses can be seen: a clear reduction of the neutrino flux at high energies takes place. 

These results indicate that the conditions in the particle acceleration region must be taken carefully into account when realistic estimates of the neutrino flux are implemented.  

\section{Perspectives}

Contrary to early optimistic predictions (e.g. White 1985), isolated massive stars are no strong sources of gamma rays. The same is valid for neutrino production. Adiabatic losses for protons in the stellar wind and inefficient acceleration are the likely causes, as pointed out long ago by V\"olk \& Forman (1982). Binary systems, instead, are promising candidates for gamma-ray and neutrino emission. These systems include colliding wind binaries and microquasars. Collective processes in star-forming regions can also result in the generation of high-energy neutrinos related to stars. Long gamma-ray bursts are interesting candidates as well, if losses in the magnetic field do not strongly attenuate the high-energy neutrino production. This would require for the interaction region for protons to be at some distance from the central source. 

Altogether we can say that the perspectives for detection of high-energy neutrinos of stellar origin by km$^{3}$-type detectors are moderately optimistic at present.     

\vspace{0.5cm}
\acknowledgements 
I congratulate Prof. Josep Mart\'{\i} for a highly stimulating meeting. I am very grateful to him, his wife and the whole LOC for their support and efforts. It has been a pleasure to participate of this event. My work on high-energy astrophysics is supported, in part, by grants AYA 2007-68034-C03-01, FEDER funds and ANPCyT (PICT-2007-00848, BID 1728/OC-AR).

\end{document}